\documentclass[12pt,preprint]{aastex}
\usepackage{psfig}

\newcommand{\be}{\begin{equation}}
\newcommand{\en}{\end{equation}}
\def\ltsima{$\; \buildrel < \over \sim \;$}
\def\lsim{\lower.5ex\hbox{\ltsima}}
\def\gtsima{$\; \buildrel > \over \sim \;$}\def\gsim{\lower.5ex\hbox{\gtsima}}

\def\ergs  {\rm \ erg \, s^{-1}}

\def\cmdue {\rm \ cm^{-2}}



\begin{document}

\title{The resolved fraction of the Cosmic X--ray Background}

\author{A.~Moretti\altaffilmark{1}, S.~Campana\altaffilmark{1}, 
D.~Lazzati\altaffilmark{2}, G.~Tagliaferri\altaffilmark{1}}

\altaffiltext{1}{INAF--Osservatorio Astronomico di Brera, Via E. Bianchi
46, Merate (LC), 23807, Italy.}
\altaffiltext{2}{Institute of Astronomy, University of Cambridge, 
Madingley Road, Cambridge CB3 0HA, UK.}

\authoremail{moretti@merate.mi.astro.it}
 
\begin{abstract}
We present the X--ray source number counts in two energy bands (0.5--2 and 
2--10 keV) from a very large source sample: we combine data of six different
surveys, both shallow wide field and deep pencil beam, performed with three
different satellites (ROSAT, Chandra and XMM--Newton).
The sample covers with good statistics the largest possible flux range so far: 
[$2.4\times10^{-17}- 10^{-11}$]  erg s$^{-1}$ cm$^{-2}$ in the soft band and
[$2.1\times10^{-16}-8\times 10^{-12}]$ erg s$^{-1}$ cm$^{-2}$ in the hard
band. 
Integrating the flux distributions over this range and taking into account the
(small) contribution of the brightest sources we derive the flux density
generated by discrete sources in both bands. 
After a critical review of the literature values of the total Cosmic X--Ray
Background (CXB) we conclude that, with the present data, the
$94.3_{-6.7}^{+7.0}\%$ and $88.8_{-6.6}^{+7.8}\%$ of the soft and hard CXB can
be ascribed to discrete source emission. 
If we extrapolate the analytical form of the Log N--Log S distribution beyond
the flux limit of our catalog in the soft band we find that the flux from
discrete sources at $\sim 3\times10^{-18}$ erg s$^{-1}$ cm$^{-2}$ is
consistent with the entire CXB, whereas in the hard band it accounts 
for only $93\%$ of the total CXB at most, hinting for a faint and obscured population
to arise at even fainter fluxes. 

\end{abstract}

\keywords{diffuse radiation -- surveys -- cosmology: observations --
X-rays: general}

\section{Introduction}

The Cosmic X--ray Background (CXB) origin and nature have attracted the
attention of astronomers since its discovery (Giacconi et al. 1962).
Diffuse emission models accounting for (a large fraction of) the CXB have
been ruled out by COBE observations (Mather et al. 1990), leaving 
the discrete faint sources hypothesis (Setti \& Woltijer 1979). 
ROSAT deep pointing on the Lockman hole region allowed to resolve about
$70-80\%$ of the CXB, at a flux level of $10^{-15}$ erg s$^{-1}$ cm$^{-2}$ in 
the soft (1--2 keV) energy band (Hasinger et al. 1998).
The great majority of sources brighter than $5 \times 10^{-15}$ erg
s$^{-1}$ cm$^{-2}$ were optically identified with unobscured (type I) AGN 
(Schmidt et al. 1998; Lehmann et al. 2001). 
Comparable results in the hard (2--10 keV) energy band have been achived only
recently thanks to Chandra and XMM--Newton. Chandra deep fields in particular
enabled us to reach limiting fluxes as low as $2\times 10^{-16}$
erg s$^{-1}$ cm$^{-2}$, resolving about $80-90\%$ of the hard CXB (Mushotzky
et al. 2000; Hornschemeier et al. 2000, 2001; Brandt et 
al. 2001; Hasinger et al. 2001; Tozzi et al. 2001; Campana 
et al. 2001; Rosati et al. 2002; Moretti et al. 2002; Miyaji \&
Griffiths 2002; Giacconi et al. 2002).
Main contributors are thought to be
absorbed and unabsorbed AGN with a mixture of quasars and narrow
emission-line galaxies as optical counterparts (e.g. Fiore et 
al. 1999; Akiyama et al. 2000; Barger et al. 2001).  

X--ray surveys can be either wide-field, covering a large area
but reaching relatively bright limiting fluxes, or pencil-beam (like the ones
performed by Chandra) over very small areas but reaching the faintest possible
flux limits. 
To our purpose, considering separately wide--field and pencil--beam surveys can
be somehow misleading: recent studies have shown that the bright and the faint
parts of the flux distributions have different slopes (e.g. Hasinger et
al. 1998 for the soft band distribution; Campana et al. 2001 for the hard
distribution). 
 From wide--field surveys it is possible to estimate 
accurately the normalization and the slope of the bright--end (Hasinger et
al. 1998 and Baldi et al. 2002 for the soft band; Cagnoni et al. 1998 and
Baldi et al. 2002 for the hard band). Many difficulties arise instead in the
calculation of the position of the break and of the faint--end slopes. In the
same way from the deepest surveys (Chandra deeps fields, Campana et
al. 2001; Rosati et al. 2002; Cowie et al. 2002; Brandt et al. 2001) 
the faint--end slope is well established, whereas due to the poor statistics of
the bright sources, the position of the break is highly uncertain.   
We compiled a single large source catalog picking up 
flux data from different (already published) surveys (both wide--field
and pencil--beam surveys).
In this way we can cover the largest flux interval so far and properly
establish the analytical form of the flux distribution.

In Section 2 we describe in some detail the surveys used in
the present analysis. 
In the soft X--ray band for the very bright part we include data from the
ROSAT-HRI Brera Multi-scale Wavelet (BMW) survey (Panzera et al. 2002)
covering the interval $10^{-14}-10^{-11}$ erg s $^{-1}$ cm$^{-2}$ with a
maximum sky--coverage of $\sim 90$ deg$^2$.  
In the very bright range of the hard band we consider the  $\sim 70$ deg$^2$ of 
the ASCA-GIS Hard Serendipitous Survey (HSS) data (Della Ceca et al. 2001;
Cagnoni et al. 1998), that covers the flux range $10^{-13}-8\times 10^{-12}$
erg s$^{-1}$ cm$^{-2}$. 
In order to fill the gap between the very bright parts and faint ends in both 
band distributions we use the HELLAS2XMM survey data (Baldi et al. 2002).
This survey has a maximum area of $\sim 3$ deg$^2$ and flux range of 
$5\times 10^{-16}-10^{-13}$ and $10^{-15}-10^{-13}$  erg s$^{-1}$ cm$^{-2}$ 
for the soft and hard band, respectively. 
Finally, as deep pencil-beam surveys we include our analysis of the Chandra Deep
Field South (CDFS, Campana et al. 2001; Moretti et al. 2002) as well as 
the Hubble Deep Field North (HDFN) analysed with the same detection 
algorithm. These two fields provide data
at the faintest end of the Log N--Log S relation: namely $2\times 10^{-17}$ erg
s$^{-1}$ cm$^{-2}$ in the soft band and $2\times 10^{-16}$ erg s$^{-1}$
cm$^{-2}$ in the hard band, respectively.  

Due to poor statistics we cut our overall distributions to $\sim 10^{-11}$ erg
s$^{-1}$ cm$^{-2}$ and $\sim 8\times 10^{-12}$ erg s$^{-1}$ cm$^{-2}$ in the soft
and hard band, respectively; in Section 3 we estimate the contribution of very
bright sources to the CXB.
In Section 4 we discuss how the presence of clusters of
galaxies in the source catalog affects our calculations. 
One of the major uncertainties involved in the estimate of the fraction of the
CXB resolved into point sources is the CXB level itself. Several estimates
have been derived with instrinsic variations of up to $20\%$ in the soft
band (1--2 keV) and up to $40\%$ in the hard (2--10 keV) band. 
A critical analysis of the CXB data is described in Section 5.
Section 6 deals with conversion factors and cross--calibration between the
different instruments. Section 7 is dedicated to the total Log N--Log S
distribution. Discussion and conclusions are reported in Section 8.

\begin{table*}[htb]
\begin{center}
\caption[]{Main characteristics of wide-field and pencil-beam surveys used to 
build the general catalog. In the fourth column the original flux limit 
values of the catalogs are reported, calculated assuming the photon
indexes reported in the fifth column (in Section 6 we describe our approach
to make all samples homogeneous).}
\footnotesize{
\begin{tabular}{l|c|c|c|c|c|c}
\
Band &Name           &Area [deg$^2$] & Limits [erg s$^{-1}$ cm$^{-2}$] & $\Gamma$ &Sources & References \\
\hline
Soft & BMW-HRI       &88.75   &$8.98\times10^{-15}-9.50\times 10^{-12}$& 2.0&3329 &Panzera et al. 2002\\
     & HELLAS2       & 2.78   &$5.89\times10^{-16}-8.44\times 10^{-13}$& 1.7&1022 &Baldi et al. 2002\\
     & BMW-CDFS      & 0.06   &$2.44\times10^{-17}-3.73\times 10^{-14}$& 1.4& 231 &Campana et al. 2001\\
     & BMW-HDFN      & 0.05   &$3.54\times10^{-17}-1.63\times 10^{-14}$& 1.4& 204 &This paper\\
\hline
\hline
Hard & ASCA-HSS      &70.82   &$1.06\times10^{-13}-7.79\times 10^{-12}$&  1.7&189 &Cagnoni et al. 1998\\
     & HELLAS2       & 2.78   &$2.81\times10^{-15}-1.04\times 10^{-12}$&  1.7&496 &Baldi et al. 2002\\
     & BMW-CDFS      & 0.06   &$2.10\times10^{-16}-8.41\times 10^{-14}$&  1.4&177 &Campana et al. 2001\\
     & BMW-HDFN      & 0.05   &$2.19\times10^{-16}-2.99\times 10^{-14}$&  1.4&164 &This paper\\

\end{tabular}}
\label{tab:sinopsi}  
\end{center}
\end{table*}

\section{The surveys} 
\subsection{BMW-HRI}
A complete analysis of the full ROSAT HRI data set with a wavelet-based detection 
algorithm (BMW-HRI) has been recently completed (Panzera et
al. 2002; see also Lazzati et al. 1999; Campana et al. 1999).
The complete catalog consists of about 29,000 sources.
From this survey, following the usual approach for serendipitous surveys
(e.g. Cagnoni et al. 1998, Baldi et al 2001), we selected high galactic latitude
fields ($|b| \geq 30^{\circ}$), with more than 5 ks exposure time, excluding the
Magellanic Clouds and Pleiades regions.
Moreover we filtered out the observations pointed on known clusters of galaxies, 
stellar clusters, supernovae remnants, Messier catalog objects and
most of the NGC catalog objects.  
Finally in the case of two or more overlapping fields we retained only the
deepest one.
All these selection criteria were applied to prevent
inclusion in the catalog of not truly serendipitous sources.
In each field we considered only sources detected in the image section  
between 3 and 15 arcmin off--axis angle: the final analysis has been carried 
out over 501 fields, corresponding to a maximum area of 88.75 deg$^2$.
The catalog consists of 3,161 sources.
The BMW-HRI distribution is very similar both in steepness and in normalization
to the bright end of the ROSAT Deep Survey (Hasinger et al. 1998), but is less
affected by cosmic variance due to the large number of fields considered.

\subsection{ASCA-HSS}
To get the bright end of the hard source distribution we took advantage of
the ASCA-HSS survey carried out by Della Ceca et
al. (2001; see also Cagnoni et al. 1998). They considered 300 ASCA GIS2 images
(at high galactic latitude, not centered on bright or extended targets,
etc.), considering the central part of the image within 20 arcmin. The sample consists of 189 serendipitous sources
with fluxes in the range $\sim 1\times 10^{-13} - 8 \times 10^{-12}$ erg
cm$^{-2}$ s$^{-1}$. The total sky area covered by the ASCA HSS is $\sim 71$
deg$^2$. 

\subsection{HELLAS2XMM}
HELLAS2XMM is a serendipitous medium-deep survey carried out on 15 XMM--Newton
fields, covering nearly 3 deg$^2$ (Baldi et al. 2002). It contains a total of 1022
and 495 sources in the soft 0.5--2 keV band and hard 2--10 keV band,
respectively. The corrisponding limiting fluxes are 
$5.9\times10^{-16}$ and $2.8\times10^{-15}$
erg s$^{-1}$ cm$^{-2}$. In the soft
band this is one of the largest samples available to date and surely the
largest in the 2--10 keV band at these limiting fluxes.

\begin{figure*} [bht]
\begin{tabular}{cc}
{\psfig{figure=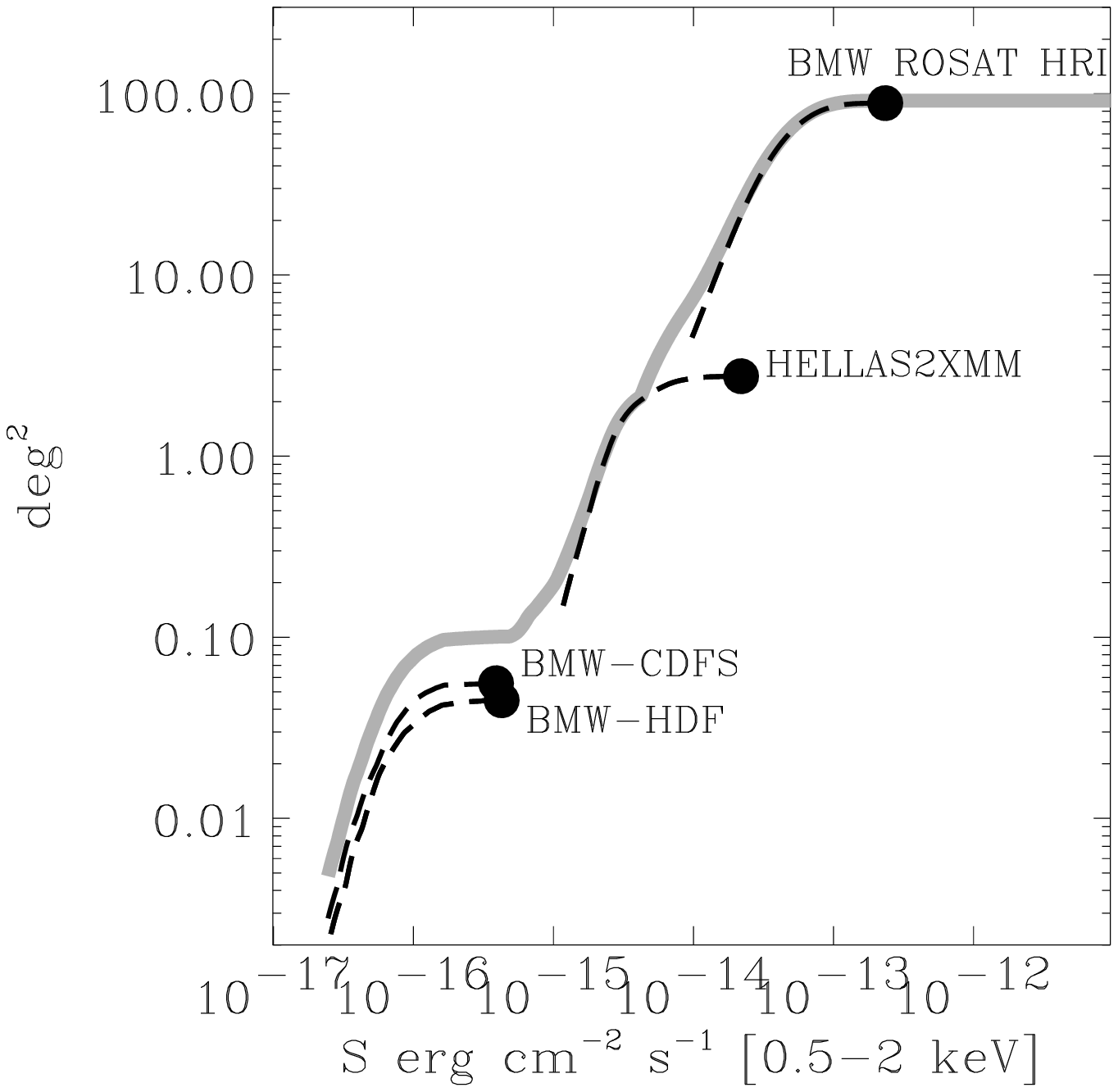,width=8cm}}&
{\psfig{figure=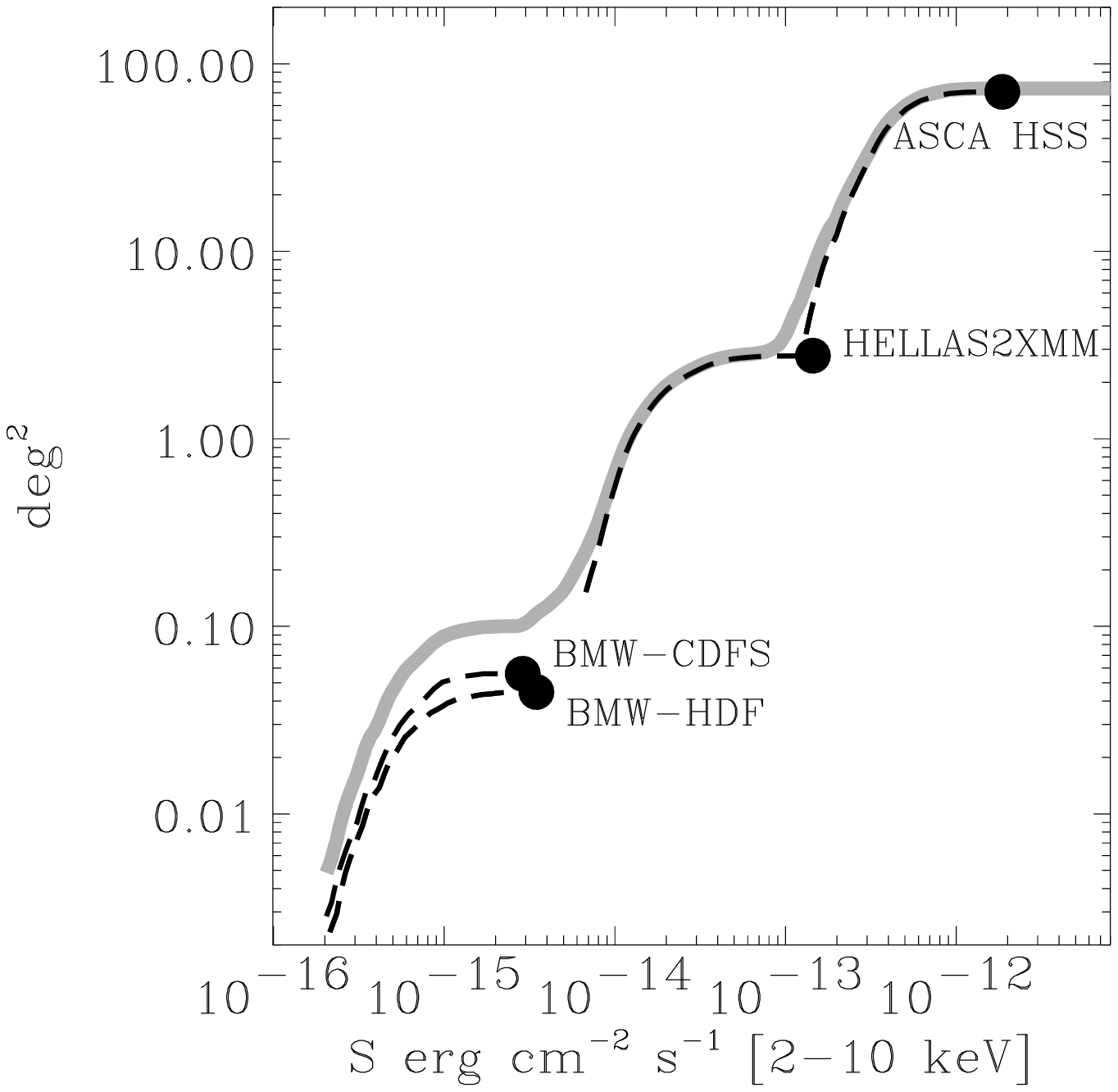,width=8cm}}\\
\end{tabular} 
\caption[]
{The sky-coverages of the the surveys described in the text are plotted.
For each survey the black dot corresponds to the minimum flux for which
the area of the survey is maximum: for brighter fluxes the sky coverage is
flat and is not plotted.
The sum is plotted with a grey line in both the soft (left panel) and hard band
(right panel). }
\label{fig:skycov} 
\end{figure*}

The sky coverage of these surveys are shown in Fig. \ref{fig:skycov}.

\subsection{Pencil beam surveys}

As pencil beam surveys we consider the two deepest look at the X--ray
sky. These were provided by the Chandra 1 Ms look at the CDFS (Rosati et
al. 2002) and at the Hubble Deep Field North (HDFN; Brandt et al. 2001). 

The CDFS consists of eleven observations for an effective exposure time of 940
ks. We analyzed the inner $8'$ radius image with a dedicated wavelet detection 
algorithm (BMW-Chandra; see Moretti et al. 2002). We detected 244 and 177
sources reaching limiting fluxes of $2.44\times 10^{-17}$ and $2.10\times
10^{-16}$ 
erg s$^{-1}$ cm$^{-2}$ in the soft (0.5--2 keV) and hard (2--10 keV)
bands, respectively. A full account of this analysis can be found in Campana
et al. (2001) and Moretti et al. (2002).

With the same detection algorithm and procedures adopted for the analysis of
the CDFS we analysed the 1 Ms exposure of the HDFN. 
The HDFN consists of twelve observations for a total nominal exposure time of
978 ks. The data were filtered to include only
the standard event grades 0, 2, 3, 4 and 6.  All hot pixels and bad columns
were removed. Time intervals during which the background rate is larger than
$3\,\sigma$ over the quiescent level were also removed in each band separately.
This results in a net exposure time of 952 ks in the soft band and 947 ks in
the hard band, respectively. 
We removed also flickering pixels with two or more events
contiguous in time (time interval of 3.2 s). 
The twelve observations were co-added with a pattern recognition
routine to within $0.4''$ r.m.s. pointing accuracy. We restricted our analysis
to the fully exposed ACIS-I area. Exposures were taken basically at two
different positions separated by $6'$. The fully exposed region thus has a
rectangular shape. Within this region we also restricted to a circular
region with $8'$ radius from the barycenter of the observations 
for sky-coverage purposes (see below). The full sky--coverage is $\sim
0.045$~deg$^2$ ($10\%$ smaller than in the CDFS).   
The average background in the considered region is 0.07 (0.12) counts s$^{-1}$
per chip in the soft (hard) band and is in very good agreement with the
expected values reported in the Chandra Observatory Guide.
We adopted a count-rate to flux conversion factors in the 0.5--2 keV and in
the 2--10 keV bands of $4.5 \times 10^{-12}$ erg s$^{-1}$ cm$^{-2}$ and of
$2.66 \times 10^{-11}$ erg s$^{-1}$ cm$^{-2}$ respectively. These numbers
were computed assuming a Galactic absorbing column of $1.6 \times 10^{20}$
cm$^{-2}$ and a power law spectrum with a photon index $\Gamma = 1.4$.

\begin{figure*} [bht]
{\psfig{figure=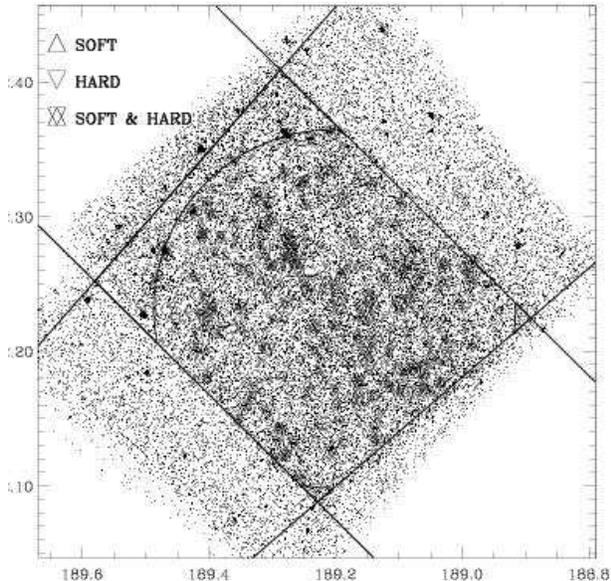,width=8cm}}
\caption[]
{The BMW surveyed area of the Chandra HDF region: we restricted our analysis 
to a circular region of 8 arcmin radius centered on the barycenter of the
maximum exposed region.
The BMW source catalog from the 1Ms Chandra observation of the HDF is the
only survey we use in this work not yet published.}
\label{fig:hdf} 
\end{figure*}

We run our BMW algorithm tailored for the analysis of Chandra fields, in 
the same way and with the same thresholds used in the analysis of the CDFS
(Campana et al. 2001; Moretti et al. 2002). 
We detected 214 and 170 sources in the soft and hard band, respectively;
39 sources ($\sim 15 \%$ of all detected sources)
are revealed only in the hard band, and 83 ($\sim 33\%$) only in the
soft band (Fig. \ref{fig:hdf}). 

As for the CDFS we carried out extensive simulations (400 fields per band) to
assess with very good accuracy the sky coverage. 
Moreover, we corrected for the Eddington bias following the approach by
Vikhlinin et al. (1995) as described in Moretti et al. (2002).
The Eddington bias starts affecting the HDFN data at a level of 
$\sim 20$ counts in the soft band ($\sim 9\times 10^{-17}\ergs\cmdue$) and  
$\sim 30$ counts in the hard band ($\sim 8\times 10^{-16}\ergs\cmdue$). 
Our simulations show that we are able to recover the number source
distribution down to 5 (7) corrected counts in the inner core of the image,
declining to 9 (11) corrected counts in the outskirts for the soft (hard)
band.   
These counts gives a flux limit in the inner region of $3.51\times 10^{-17}$ 
erg s$^{-1}$ cm$^{-2}$ and $2.29\times10^{-16}$ erg s$^{-1}$ 
cm$^{-2}$ in the soft and hard band, respectively.
The sky coverage of these surveys are shown in Fig. \ref{fig:skycov}. 

In order to evaluate the possible cosmic variance between the two deep 
fields we compared the faint end  of the two flux distributions.
In both cases we found that, excluding the bright sources 
($> 5\times 10^{-15}\ergs\cmdue$), the values of the analytical fits (slope and 
normalization) realtive to the two fields are compatible at $1 \sigma$ level
to each other and with the values relative to the fit of the entire sample
(see below).  
    
\section{Contribution of very bright sources to the background}

The surveys under consideration lack of a proper covering at the brightest
flux levels: in fact most of the X--ray brightest sources are the target of the
observation and are excluded from serendipitous catalogs.
For this reason in these surveys we cut the source flux distributions at
$10^{-11}$ ($8\times 10^{-12}$) erg cm$^{-2}$ s$^{-1}$ in the soft (hard)
band. Note that even if the number of these bright sources is relatively
small (less than a few hundred sources on the whole sky), 
their contribution to the CXB is not negligible. 

To overcome this problem, in the case of the soft band, we took advantage
of the ROSAT Bright Survey
(RBS, Schwope et al. 2000) that contains all sources of the ROSAT All Sky 
Survey (RASS) sources with count rate larger than 0.5 c s$^{-1}$.
There are 93 high galactic latitute extragalactic sources brighter 
than $10^{-11}$ erg
cm$^{-2}$ s$^{-1}$ which provide a density flux of 
$F_{S11} = 1.45\times 10^{-13}$ erg cm
$^{-2}$ s$^{-1}$ deg$^{-2}$ in the 1--2 keV energy band or $3.2\%$ of the CXB
(see below).   

For the hard band we considered the HEAO-1 A2 extragalactic survey which is  
complete down to $3.1\times 10^{-11}$ erg cm$^{-2}$ s$^{-1}$ (Piccinotti et
al. 1982). The total flux of the 66 sources amounts to $F_{H11} = 
4.28 \times 10^{-13}$ erg cm$^{-2}$ s$^{-1}$ deg$^{-2}$ or $2.1\%$ of the CXB
(see below). 
To include the small intermediate interval ($8.0\times 10^{-12}$ and $3.1\times
10^{-11}$ erg cm$^{-2}$ s$^{-1}$) not covered by the hard X--ray surveys
we will extrapolate the Log N--Log S relation.

\section{Contribution from extended sources}

A significant fraction of the CXB is made up by the thermal
bremsstrahlung emission from clusters of galaxies.
In the soft band (1--2 keV) this fraction is estimated at a level of $\sim 6
\%$ from direct measurments of the cluster Log N--Log S 
(Rosati et al. 1998, 2002). In the
hard band there is not a precise determination, but this can be
estimated to be at a level of $\sim 5 \%$ from the Log N--Log S (e.g. 
Gilli et al. 1999, derived from Ebeling et al. 1997). 
So far, in the building of the general source catalog, no selection
have been made among different kind of sources. Clusters of galaxies
are included in our catalog as well the other cosmological sources
(AGN and QSO). 

In the construction of the Log N--Log S, if we treat clusters of galaxies in
the same way as point-like sources, we then introduce flaws.
First, the X--ray spectrum of a cluster of galaxies is different from the
other point-like sources and therefore the conversion factor changes.
More importantly, clusters of galaxies, having extended emission, have different
sky--coverages with respect to point--like sources 
(for a given instrument, at the same flux, in general, a
point--like source is more easily detected than an extended 
one). Thus, if we use the point-like sky coverage for all sources,   
we underestimate the level of the integrated flux because we underestimate
the statistical weight of the extended sources.
Clearly, this difference is particulary pronunced in the case of surveys based
on high spatial resolution instruments (Chandra, ROSAT-HRI, XMM--Newton),
whereas it is neglegible for the ASCA-HSS survey.

To evaluate the contribution of the extended sources to the CXB
we can use only our BMW surveys
(i.e. ROSAT-HRI and Chandra), for which we have an extension flag.
In the BMW-HRI catalog we have 199 extended sources, which correspond 
to $5\%$ of the sources of the catalog.
An appropriate sky coverage for the extended sources of the BMW-HRI catalog has
been derived from extensive Montecarlo simulations described in Moretti et
al. (2003). We estimate that the contribution of these clusters treated
as extended and not point--like, increases, going from $3.5\%$ to $\sim 6\%$
of the total 1-2 keV CXB flux (see next Section) 
which is in very good agreement with Rosati et al. (2002).
At lower fluxes we estimate an extra contribution from clusters in the
HELLAS2XMM survey of $\sim 1\%$. Being this just an estimate we include it
in the error budget.

In the hard band, the arcmin angular resolution of ASCA makes any correction
due to the presence of extended sources negligible. We estimate, from 
preliminary classification of HSS sources (Della Ceca et al. 2001), that the
total correction due to the changing in the conversion factor is $< 1\%$ and 
we include it in the calculation of the uncertainties. 
At fainter fluxes, the
contribution of extended sources in the hard band of the HELLAS2XMM survey is
estimated to be $<1\%$ which is again included in the error budget.   
 
\begin{figure*} [bht]
{\psfig{figure=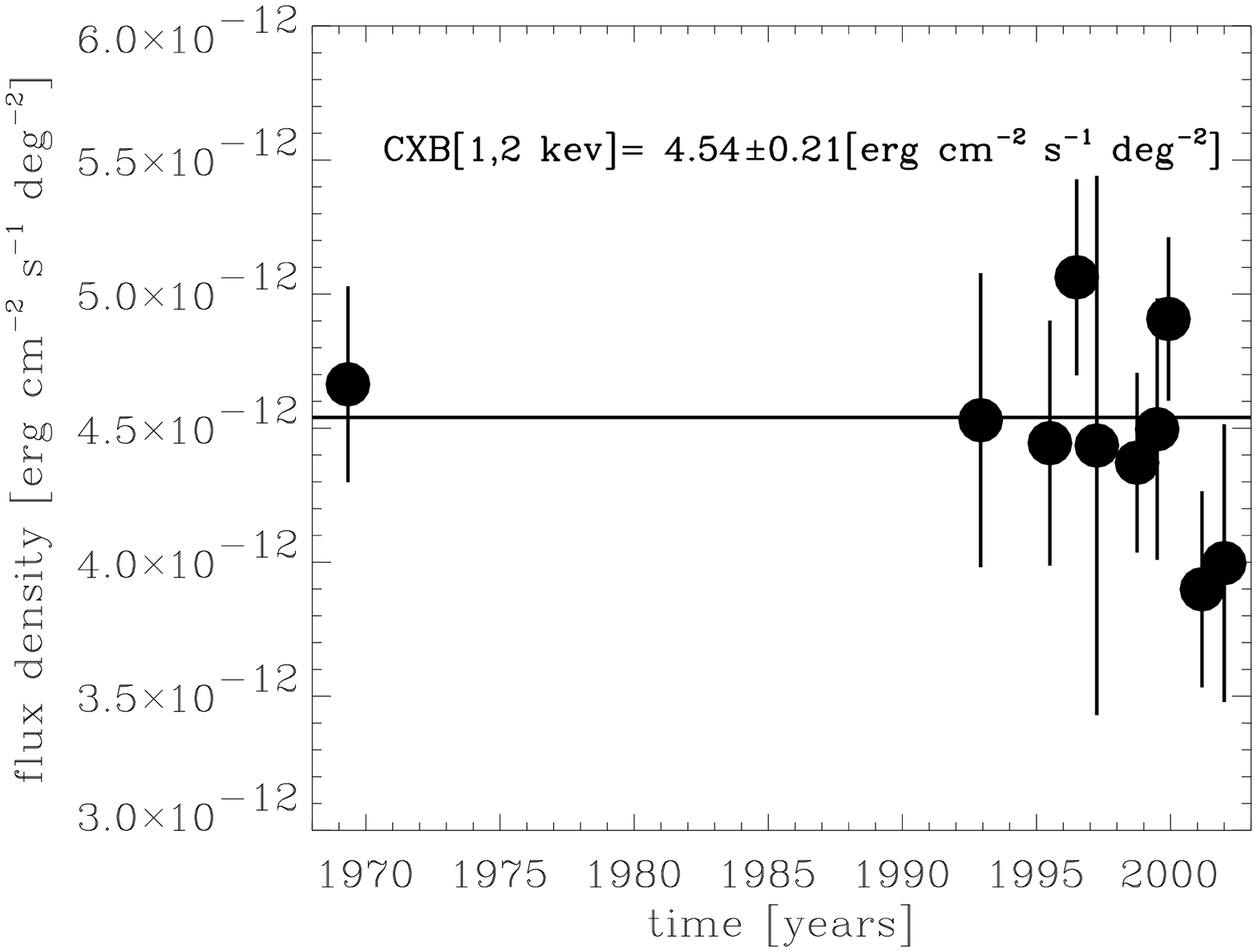,width=8cm}}
{\psfig{figure=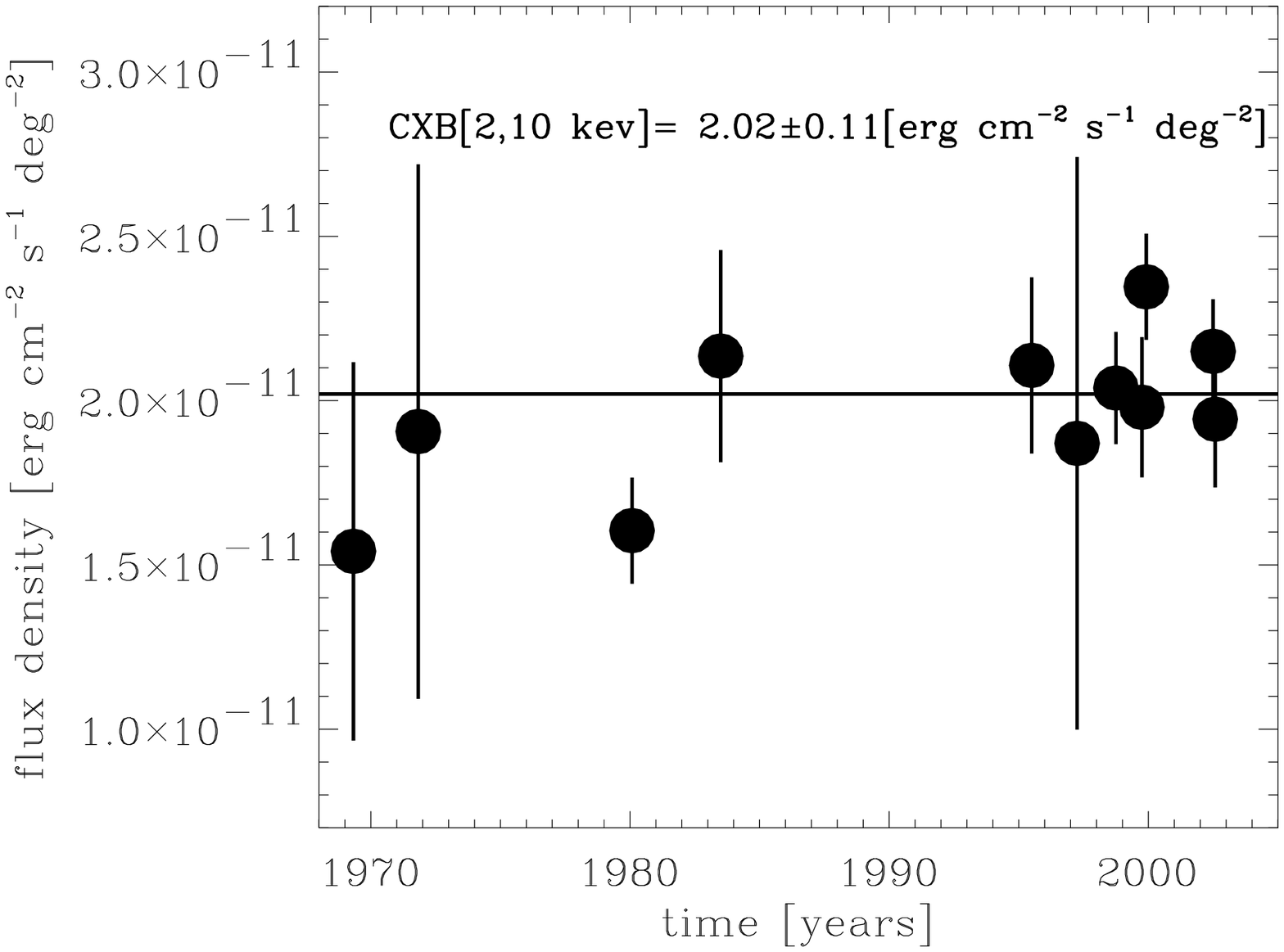,width=8cm}}
\caption[]{{\bf Left panel}: soft CXB (1--2 keV) measurements spaced in time along with their
$68\%$ error bars; the
constant line represents the best fit to all these measurements
($\chi^2_{\rm red}=0.9$).
CXB values (left to right) are from Gorenstein et al. (1969); Garmire et al. (1992);
Gendrau et al. (1995); Georgantopoulos et al. (1996); Chen et al. (1997);
Miyaji et al. (1998); Parmar et al. (1999); Vecchi et al. (1999); Kuntz et
al. (2001); Lumb et al. (2002). 
{\bf Right panel}: hard CXB (2--10 keV) measurements spaced in time  along with their
$68\%$ error bars;
the constant line represents the best fit to all these measurements 
($\chi^2_{\rm red}=1.3$).
CXB values (left to right) are from Gorenstein et al. (1969); Palmieri et al. (1971);
Marshall et al. (1980); McCammon et al. (1983); Gendrau et al. (1995); Chen et
al. (1997); Miyaji et al. (1998); Ueda et al. (1999); Vecchi et al. (1999);
Lumb et al. (2002); Kushino et al. (2002).}
\label{fig:bck}
\end{figure*}

\section{X--ray background level}

Measuring the CXB flux has been one of the most challanging tasks of X--ray
astronomy. Several measurements have been carried out with rockets and
satellites.
Barcons et al. (2000) found that, while the differences in the measurments
among different studies using different data from the same instrument can be ascribed 
to the cosmic variance, systematic differences remain among different 
missions. 
Following Gilli (2002), we made a bibliographic search selecting
ten and eleven measurements in the soft (1--2 keV) and hard (2--10 keV) energy
bands, respectively.  
We compute $68\%$ errors estimates on the flux adding the
contribution of the bright sources when they have been excluded from the
analysis (Hasinger 1996). Our results are reported in Fig. \ref{fig:bck}. A
fit with a constant provides a good representation in terms of reduced
$\chi^2$. In the soft band (1--2 keV) we derive a value of
$(4.54\pm0.21)\times 10^{-12}$ erg s$^{-1}$ cm$^{-2}$ deg$^{-2}$ ($90\%$
confidence level) with a $\chi^2_{\rm red}=0.9$. Assuming the average value
for the slope of the spectrum ($\Gamma=1.4$, Rosati et al. 2002) this value
correspond to $(7.53\pm0.35)\times 10^{-12}$ erg s$^{-1}$ cm$^{-2}$ deg$^{-2}$
in the 0.5--2 keV band. In the hard band we obtain $(2.02\pm0.11)\times
10^{-11}$ erg s$^{-1}$ cm$^{-2}$ deg$^{-2}$ with a $\chi^2_{\rm red}=1.3$
($32\%$ null hypothesis probability).
These values, in both band, are in excellent agreement with the CXB intensity 
value reported in Barcons et al. (2000).
Thus, despite the variability reported in the literature our fit indicates
that the different estimates of the soft and hard CXB are consistent each
other in a statistical sense.

\section{Conversion factors and cross--calibrations}

The CXB is the result of the integrated emission of a mix of different
sources, mainly unabsorbed and absorbed AGNs. The resulting spectrum in the energy 
range of our interest can be modelled with a power law with $Gamma \sim 1.4$,
which is very different from the spectrum slope of the typical sources which
make it: this is the so called spectral paradox and was explained for the
first time by Setti \& Woltijer (1979).
An important point in our work is the choice of the spectrum for converting
counts to fluxes.
In all the surveys we used to build the catalog a single spectrum slope
$\Gamma$ has been assumed; these values, reported in Table \ref{tab:sinopsi},
have been chosen to match the expected average spectrum of the sources:
they change significantly from survey to survey passing from
$\Gamma=2$  in the BMW to $\Gamma=1.4$ in the Chandra deep surveys
(Table \ref{tab:sinopsi}).  
Cowie et al. (2002) in the Chandra deep field analysis adopted $\Gamma=1.2$
pointing out that the average spectra at very faint fluxes 
is harder than that of the CXB.
In different ranges of flux the slope of the average spectrum revealed from
the staked spectrum analysis of the sources change passing from steeper values 
to shallower values with the flux lowering (e.g. Tozzi 2001). 
In the case of our work we have two requirments: the first is to make all the
sample homogeneous and the second is that because we use a very large
flux range we have to account for  the changing of the CXB spectrum as the 
flux lowers.
Our approach is the following: using literature data we attributed to each
source a different spectral slope as a function of its flux.
For this reason we collected spectral indexes from bright to faint ends from 
several surveys (soft: Vikhlinin et al. 1995; Brandt et al. 2001; 
hard: Della Ceca et al. 1999; Rosati et al. 2002).
These power law indexes have been fitted (with a Fermi function) as a function
of the X--ray flux (in the soft and hard band separately). 
Then we checked that the integrated spectrum is consistent with the expected one:
in both bands we considered the integrated spectrum built summing the
contribution of each source weighted by the sky coverage and we found that
they are both perfectly consistent with the expected one ($\Gamma~1.4$), 
having taken into account also the contribution of the clusters.
The flux of each source has then been corrected for the ratio of the nominal 
conversion factor used in the survey and the one recalculated by us, with the
interpolated power law index at the appropriate column density.
Flux corrections for single sources are on average $\sim 5\%$ 
($\sim 7\%$ in the hard band) and always less than $17\%$. 

Absolute cross--calibration between XMM--Newton and Chandra have not been yet
well explored. Lumb et al. (2001) found that Chandra fluxes are sistematically
$10 \%$ higher than XMM ones, once the differences of the detection procedures
have been took into account, without any trend with spectral slope, off--axis
angle or brightness.
In order to evaluate how the different normalization of the different
instruments could affect our calculation, we artificially increased and reduced
the flux of the single survey (one by one) by a $10\%$ factor (modifying
the corresponding sky--coverage). We found that we have typical differences of
$2\%$ of the total CXB (and never larger than $3\%$).

\section{Global Log N--Log S}

The cumulative flux distribution (Log N--Log S) at each flux $S$ 
is the number of all sources brighter than $S$ weighted by the 
corresponding sky--coverage:
\be
N(S) = \sum_{S_i > S} \frac{1}{\Omega_{\rm tot}(S_i)~ }  
\en
here the sky--coverage $\Omega_{\rm tot}$ is the sum of the
contributions of all surveys in each band (Fig. \ref{fig:skycov}).  
Given the large flux interval spanned, we consider as analytical form of
the integral source flux distribution two power laws with 
index $\alpha_{\rm 1,s(h)}$ and $\alpha_{\rm 2,s(h)}$ for the bright and faint
part respectively, joining without discontinuites at the flux $S_{\rm 0,s(h)}$: 
\begin{equation}
N(>S) = N_{S(H)}\, \Bigl[
{{(2\times 10^{-15})^{\alpha_{1,s(h)}}} \over
{S^{\alpha_{1,s(h)}}+S_{\rm 0,\,s}^{\alpha_{1,s(h)}-\alpha_{2,s(h)}}\,
S^{\alpha_{2,s(h)}}}} \Bigr] \ {\rm cgs}.
\end{equation}

To fit the data we applied a maximum--likelihood algorithm to the differential
flux distribution corrected by the sky coverage 
\be
\frac{dN}{dS} \times \Omega(S) \ .
\en
Once we obtain the analytical form of the flux distribution we can calculate
the total contribution of sources, $F_{\rm sou}$, to the CXB by integrating
the quantity 
\be
F_{\rm sou} = \int_{S_{\rm min}}^{S_{\rm max}} \Big(\frac{dN}{dS} \Big)
\times S ~ dS 
\en
with $S_{\rm min}$ and $S_{\rm max}$ as the boundary fluxes of our interval of
interest. 

\begin{figure*} [bht]
\begin{tabular}{c}
\psfig{figure=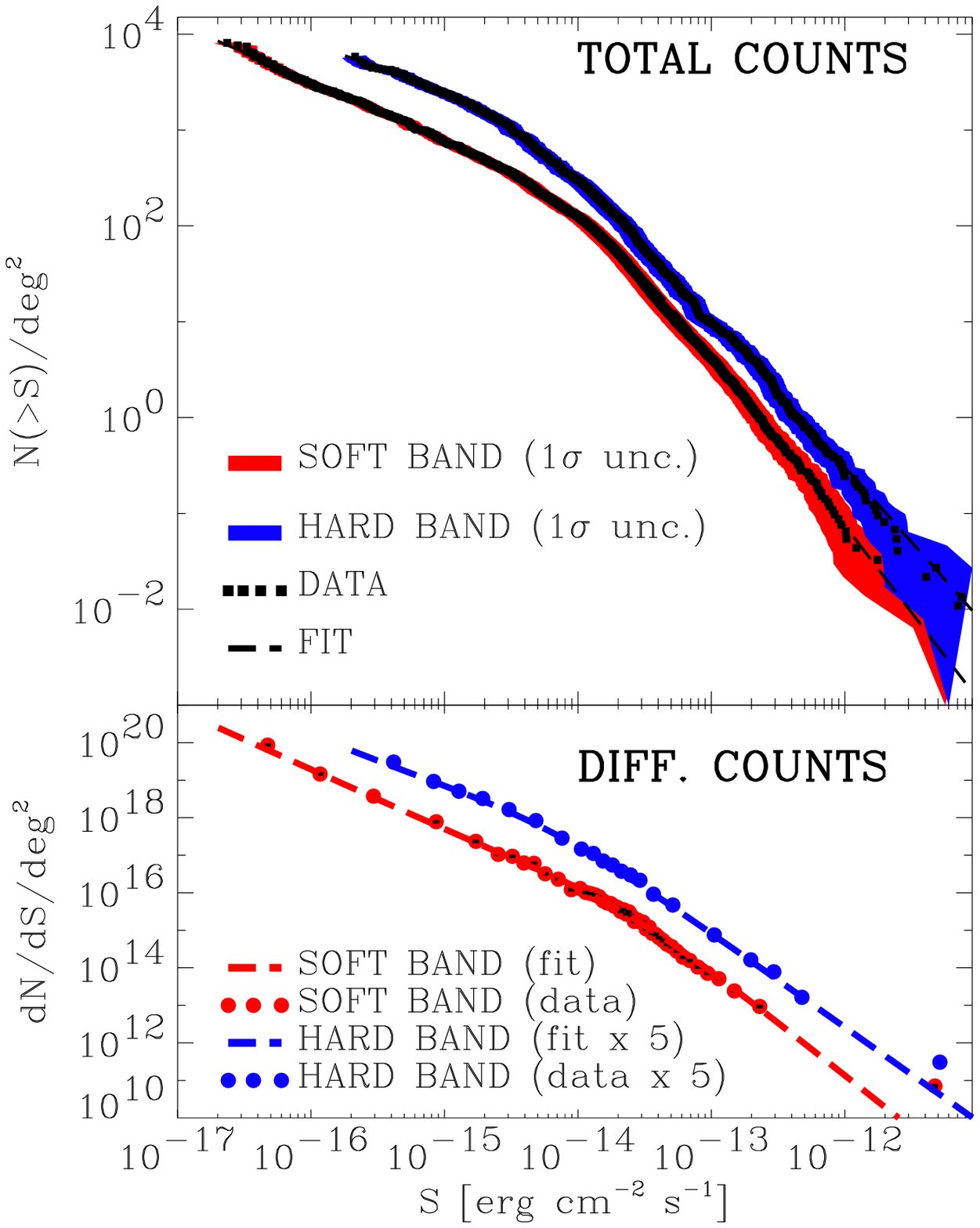,width=11cm}\\
\end{tabular}
\caption[]{In the {\bf upper panel} the Log N--Log S(integral) curves of our composite
catalog in both soft and hard band are shown: in both cases we obtained an excellent 
fit with two smoothly joined power laws (see text). 

In the {\bf lower panel} we plot the differential
distributions: the data are grouped to have a minimum of 100 (50) measures in
each bin in the soft (hard) band. The hard band counts are multiplied by 5 for
clarity of the plot.
Due to the very large y-axis range the error bars
are not visible in the graph: they are approximately 
10$\%$ and 15$\%$ respectively.}
\label{fig:glob_lnls}
\end{figure*}

\subsection{Soft energy band}

The three source distributions (BMW-Chandra, XMM2HELLAS and BMW-HRI) containing
point and extended sources, cover with good signal to noise ratio the flux range
$2.4\times 10^{-17}-10^{-11}$ erg cm$^{-2}$ s$^{-1}$.
We found a good fit with $\alpha_{\rm 1,s}=1.82^{+0.07}_{-0.09}$, $\alpha_{\rm
2,s}=0.60^{+0.02}_{-0.03}$, $S_{\rm 0,s}=(1.48^{+0.27}_{-0.31}) \times
10^{-14}$ erg cm$^{-2}$ s$^{-1}$ and $N_S=6150^{+1800}_{-1650}$ (errors at
$68\%$ confidence for the four parameters i.e. $\Delta \chi^2=4.72$, see
Fig. \ref{fig:glob_lnls}).

In order to calculate the reduced $\chi^2$ we adaptively binned the data
to contain 50 sources per bin: we obtain $\chi^{2}_{red}=1.4$ 
with 87 degrees of freedom (null hypothesis probability $\sim 1\%$). 
From the residual analysis we find that the largest scatter between data
and fit is in the knee region, where the two power laws join.
This has to be ascribed to the choice of the analitycal function
rather than a mismatch between different surveys: a function with one more 
parameter could improve the reduced $\chi^2$ value. 

As expected, the slope of the bright part is consistent with the previous
determinations. Panzera et al. (2002) has already shown
that the BMW HRI log N--log S is in excellent agreement with the 
bright part of the distribution reported by Hasinger et al. (1998).
Here, using also the HELLAS2XMM data, we find a slightly steeper value for the
bright slope (consistent at $1 \sigma $ level): $1.68 \pm 0.27 $ vs. $1.82 \pm 0.09$.
In the faint end we find good agreement with Rosati et al. (2002), 
Brandt et al. (2001) and Mushotzky et al. (2000) who report $ 0.60\pm 0.13$, 
$0.6\pm 0.1$ and $0.7\pm 0.2$, respectively.

The fitted Log N--Log S distribution gives in the 0.5--2.0 keV band an integrated 
flux $F_{\rm sou} = 6.85^{+0.28}_{-0.23} \times 10^{-12}$ erg s$^{-1}$
cm$^{-2}$ deg$^{-2}$.
This corresponds to $91.0^{+3.8}_{-3.1}\%$ of the corresponding CXB value. 
Adding the contribution at brighter fluxes (see Section 3) and  taking into account
the uncertainties in the correction for XMM2HELLAS clusters of galaxies
(Section 4) and the  uncertainties in the conversion factor and in the
cross--calibration (Section 6) we end up with 
$F_{S11}+F_{\rm sou}=94.3_{-6.7}^{+7.0}\%$
of resolved CXB (see Fig. \ref{fig:fralim}).

\subsection{Hard energy band}

We fit with the same functional form of equation (4) the hard Log N--Log S
distribution. 

The flux interval with good signal to noise ratio is $2.1\times 10^{-16}-8.0\times
10^{-12}$ erg cm$^{-2}$ s$^{-1}$.
We found a good fit with $\alpha_{\rm 1,h}=1.57^{+0.10}_{-0.08}$, $\alpha_{\rm
2,h}=0.44^{+0.12}_{-0.13}$, $S_{\rm 0,h}=(4.5^{+3.7}_{-1.7}) \times
10^{-15}$ erg cm$^{-2}$ s$^{-1}$ and $N_H=5300^{+2850}_{-1400}$ (errors at
$68\%$ confidence as before, see Fig. \ref{fig:glob_lnls}).

In order to calculate the reduced $\chi^2$ we adaptively binned the data
to contain 25 sources per bin: we obtain $\chi^2_{red}=0.93$ 
with 38 degrees of freedom (null hypothesis probability $\sim 60\%$). 
This assures us of the goodness of the fit and the effective possibility to
smoothly match data from different surveys performed with different
instruments also in the hard band.

In the bright part, after summing the HELLAS2XMM data to the ASCA--HSS data,
we find a slightly steeper value (still consistent at $1 \sigma$ level) with
respect to the value reported in Cagnoni et al. (1998) and the one based on
BeppoSAX (Giommi, Perri \& Fiore 2000).
Our determination of the faint hard slope
($\alpha_{\rm2,h}=0.44^{+0.12}_{-0.13}$) is flatter and only marginally
consistent with Rosati et al. (2002) (0.61$\pm 0.10$), Cowie et al. (2002)
(0.63$\pm 0.05$) and Mushotzky et al. (2000) (1.05$\pm 0.35$ using a single
power law).  
This is probably correlated to the different position that we estimate for
the knee of the double power law ($S_{\rm 0,h}=(4.5^{+3.7}_{-1.7})
\times 10^{-15}$ erg cm$^{-2}$ s$^{-1}$) which is fainter both than the
value reported in Cowie et al. (2002) ($1.2 \times 10^{-14}$) erg cm$^{-2}$
s$^{-1}$  and the value reported in Rosati et al. (2002)  ($ \sim 8\times
10^{-15}$) erg cm$^{-2}$ s$^{-1}$. 

Our fitted Log N--Log S distribution gives an integrated flux 
$F_{\rm sou}= 1.75^{+0.11}_{-0.10} \times 10^{-11}$ erg s$^{-1}$ cm$^{-2}$ deg$^{-2}$. 
This accounts for $86.7^{+5.5}_{-5.9}\%$ of the CXB 
(the $1\,\sigma$ uncertainties interval are reported). 
Adding the contribution of brighter sources (Section 3)
and taking into account the uncertainties in the
conversion factor and in the cross--calibrations (Section 6) and the 
uncertainty of the contribution of extended sources (Section 4) we obtain
a resolved fraction of
$F_{H11}+F_{\rm sou}=88.8_{-6.6}^{+7.8}\%$ (see Fig. \ref{fig:fralim}).

\section{Summary and conclusions}

\begin{figure*} [bht]
\begin{tabular}{cc}
{\psfig{figure=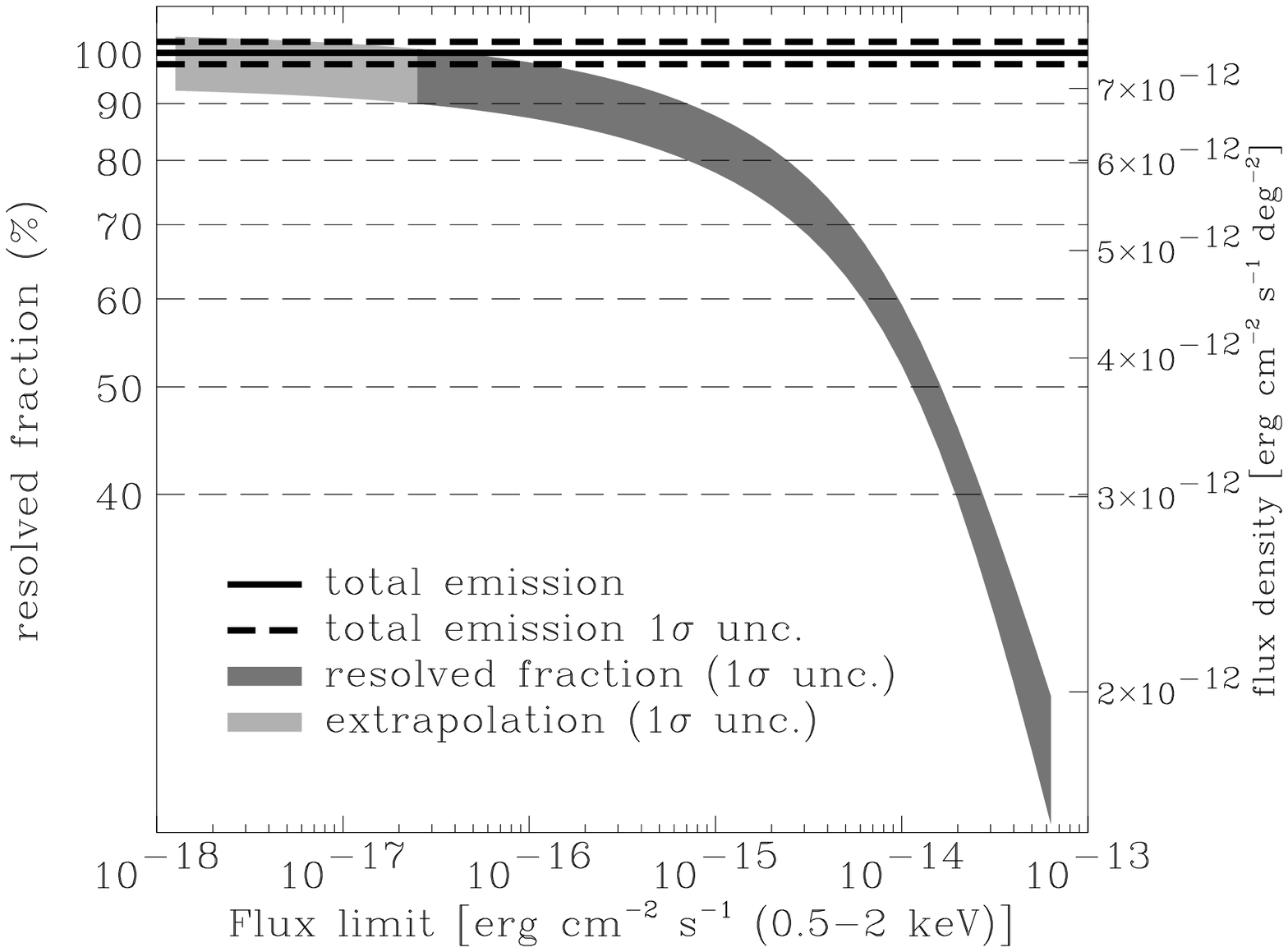,width=8cm}}&{\psfig{figure=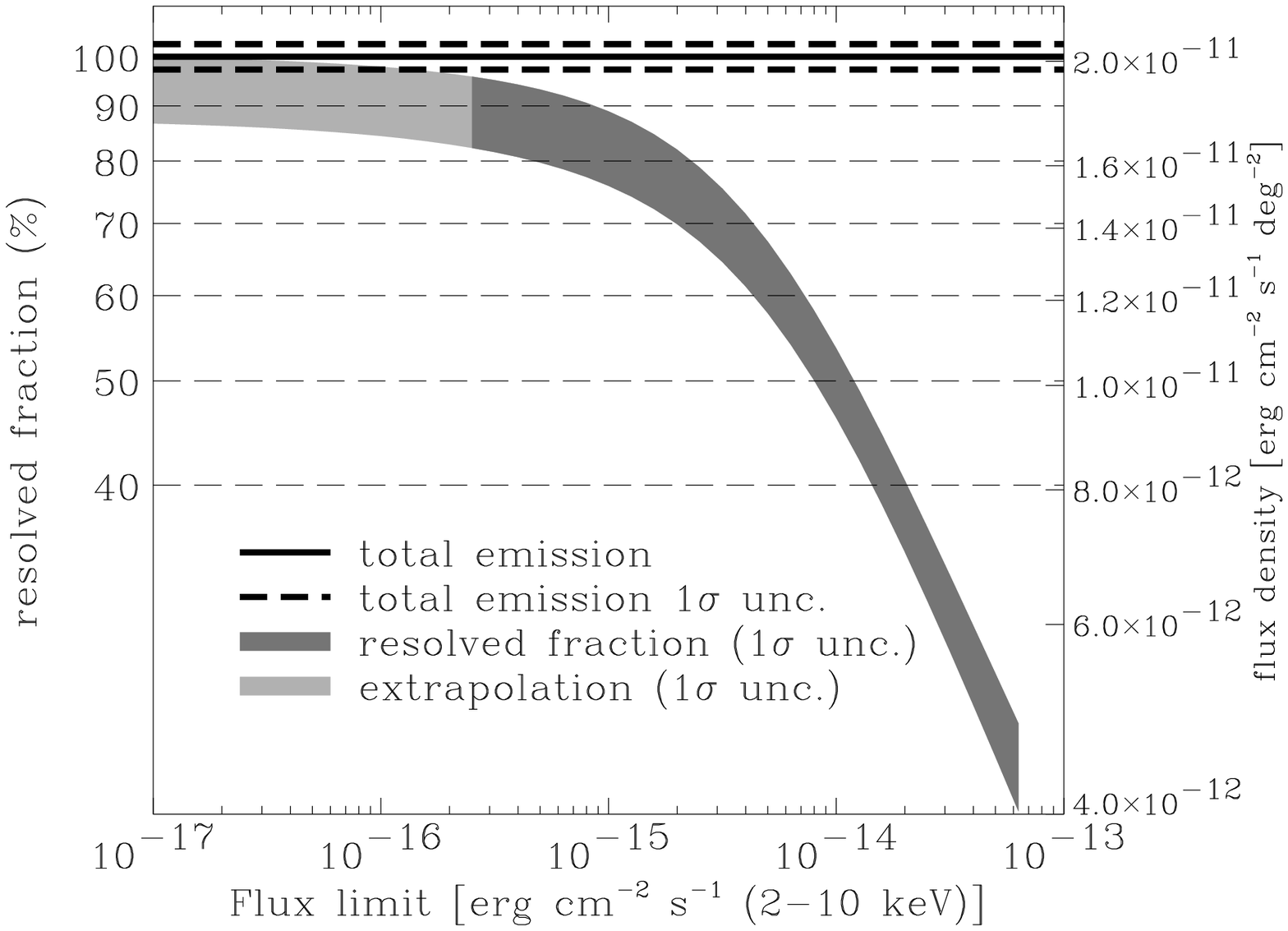,width=8cm}}\\
\end{tabular} 
\caption[]
{The fraction of resolved background as function of the flux limit in
the soft 0.5--2 keV (upper) and hard 2--10 keV (lower) energy bands.
On the right axis of each plot the absolute value of the flux density
is reported in the 0.5--2 and 2--10 keV}
\label{fig:fralim} 
\end{figure*}

While the high spatial resolution and positional accuracy of the Chandra satellite
allow us to investigate the faintest sources which make up the CXB and
to identify most of the detected photons as emission from discrete sources,
the fluctuations in the source number counts among the different Chandra
deep fields reach a $30\%$ level for the bright sources (see Setion 2.4 and Tozzi 2001).
These fluctuations correspond to very high uncertainties in the calculation of
the fraction of the total CXB that we can resolve in discrete sources.
In order to improve the statistics we matched data from different
surveys. The resulting composite catalog allowed us
to draw with good statistics the Log N--Log S curve over the maximum 
possible flux range with the data currently available:
[$2.44\times10^{-17}-1.00\times 10^{-11}$]  erg s$^{-1}$ cm$^{-2}$ in the soft
band and [$2.10\times10^{-16}-7.79\times 10^{-12}]$ in the hard band
(see Fig. \ref{fig:glob_lnls}).
Moreover, we derived a reliable value of the measure of the total CXB
by means of a critical review of the literature values.
We calculated that in the range of our composite catalogs the detected sources 
make up $94\%$ and $89\%$ of  the total soft and 
hard CXB emission, respectively. 
We obtained a good fit of the flux distribution in both bands
with two smoothly joined power laws: this demonstrates that we can use 
data obtained with different instruments in a coherent manner.

If we extrapolate the analytical form of the Log N--Log S distribution 
beyond the flux limit of our catalog in the soft band we find that 
the integrated flux from discrete sources at $\sim 3\times10^{-18}$ 
erg s$^{-1}$ cm$^{-2}$ (a factor 10 lower than the catalog limit)
is $96\%$ of the total CXB and it is consistent with its full value at $1 \sigma$
level (comparing the best value for the integrated flux from discrete sources
with the CXB $1 \sigma$ uncertainty).

In the hard band, extending again to lower fluxes the Log N--Log S 
distribution we can make up only $93\%$ of the total CXB, at most.
This is only marginally consistent with the CXB total value.
The small contribution of the faint sources is due to the fact that 
the Log N--Log S distribution converges less than logarithmically 
(Fig. \ref{fig:fralim}). 
This leaves space to the presence of a class of very faint hard sources
only poorly detected in the 2--10 keV band within the actual limits or even to 
diffuse emission. 
This class of sources could consist of heavily absorbed AGN which are expected
to provide higher contributions to the X--ray counts at higher energies. 
A possible indication of the existence of this population could be the
steepeer source counts found in the very hard band (5--10 keV), as reported 
in Rosati et al. (2002).
According to the model by Franceschini et al. (2002, see also Gandhi \& Fabian 
2002), which is based on IR statistics, in the 2--10 keV band the contribution
of obscured AGN would become dominant at $\sim 10^{-15}$ erg s$^{-1}$ cm$^{-2}$.
A qualitative study of this model allow us to estimate that the existence of
such a class of sources would result in a steepening of the hard Log N--Log S below 
$\sim 4\times10^{-16}$ erg s$^{-1}$ cm$^{-2}$ which could fill the remaining
fraction of unresolved CXB. The approximate extra--contribution is estimated 
to be about $\sim 10^{-12}$ erg s$^{-1}$ cm$^{-2}$deg$^{-2}$ (5$\%$ of the total) in the range
between $4\times10^{-17}$ and $2\times 10^{-16}$ erg s$^{-1}$ cm$^{-2}$.
Actually our data could neither confirm nor reject this eventual steepening
being very close to the limit of our catalog. 
Another possibility recently put forward is represented by X--ray emission
from star-forming galaxies that can make up to $11\%$ of the hard CXB by
extrapolating the radio counts down to 1 Jy or $10^{-18}$ erg s$^{-1}$
cm$^{-2}$ in the soft X-ray band (Ranalli et al. 2002). 
In this contest the analysis of the HDF Chandra deeper observation (2Ms) and
the XMM--Newton surveys will be crucial.


\begin{acknowledgements}
We thank Alessandro Baldi, Silvano Molendi and all the HELLAS team for
supplying data of the HELLAS2XMM survey.
We thank Silvano Molendi also for his useful comments and discussion.
We thank Roberto Della Ceca and Ilaria Cagnoni for supplying data of ASCA--HSS
survey. This work was partially supported through CNAA, Co-fin, ASI grants and Funds for Young
Researchers of the Universit\`a degli studi di Milano. 

\end{acknowledgements}

\clearpage
\end{document}